\documentclass[a4paper,10pt]{article}
\usepackage[english]{babel}
\usepackage[T1]{fontenc}
\usepackage[utf8]{inputenc}
\usepackage{amssymb}
\usepackage{physics}
\usepackage{amsmath}
\usepackage{graphicx}
\usepackage[unicode]{hyperref}
\usepackage{tikz}
\usepackage{color}
\usepackage{genyoungtabtikz}
\usepackage[title]{appendix}
\usepackage{indentfirst}
\usepackage{bm}
\usepackage{xhfill}
\usepackage{bbm}
\usepackage{multirow}
\usepackage{array}
\usepackage[centertableaux]{ytableau}
\usepackage{multicol}
\usepackage[shortlabels]{enumitem}
\usepackage{forest}
\usepackage{float}

\usetikzlibrary{decorations.pathreplacing}

\usepackage{ytableau}

\hypersetup{
    colorlinks=true,
    linkcolor=blue,
    filecolor=magenta,      
    urlcolor=cyan,
}
 
\usepackage{amsthm}

\theoremstyle{definition}

\usepackage[nottoc]{tocbibind}

\usepackage{authblk}

\title{\textbf{On palindromic numerators of bigraded symmetric orbifold Hilbert series and Kostka-Foulkes polynomials}}

\author[1,2]{\textbf{Yannick Mvondo-She}}
\affil[1]{School of Physics and Mandelstam Institute for Theoretical Physics,
University of the Witwatersrand, Johannesburg, Wits, 2050, South Africa}
\affil[2]{National Institute of Theoretical and Computational Sciences, Private Bag X1, Matieland, South Africa}
\affil[ ]{\texttt{yannick.mvondo-she@nithecs.ac.za}}

\date{}

\begin{document}

\maketitle

\begin{abstract}
From our work on partition functions in log gravity, we show that the palindromic numerators in two variables of bigraded symmetric orbifold Hilbert series take the form of sums of products of Kostka-Foulkes polynomials associated with a pair of partition $\lambda$ and $\mu=(1^n)$. The log partition function also being a KP $\tau$-function, our work gives a new description of Hall-Littlewood and Kostka-Foulkes polynomials as palindromic numerators of quotient expansions in the moduli space of formal power series solutions of the KP hierarchy. Using the structure and properties of the log partition function, we also show that the palindromic polynomials are eigenvalues of a differential operator arising from a recurrence relation and acting on the Hilbert series.
\end{abstract}

\tableofcontents

\section{Introduction}
In \cite{Mvondo-She:2018htn}, we introduced and exploited properties of counting formulae for the 1-loop graviton partition function of log gravity 
\cite{Grumiller:2008qz,Grumiller:2013at} on the thermal $\text{AdS}_3$ background originally calculated in \cite{Gaberdiel:2010xv}, with the aim of providing a unifying interpretation for both single and multiparticle log sectors. It was shown that the multi-particle states are simply built as a Fock space from products of the single particle states, by reformulating the log partition function in terms of the plethystic exponential (PE), which defines the $n^{\text{th}}$ symmetric product of functions.

Subsequently, we showed in \cite{Mvondo-She:2019vbx} that the moduli space of the log sector is the $S^n \left( \mathbb{C}^{2} \right)$ Calabi-Yau orbifold. The modus operandi was to show that the log partition function is a generating function of Hilbert series of the polynomial ring $\mathbb{C} \left[ x_1, \ldots, x_n ,y_1 ,\ldots, y_n \right]^{S_n}$ which is Cohen-Macaulay, and that the numerators of the Hilbert series are palindromic, therefore indicating that the ring being enumerated is Gorenstein, and thus establishing the Calabi-Yau nature of the log sector's moduli space.

As an expansion in terms of Schur polynomials, i.e Schur symmetric functions expressed as polynomials of power sums, in \cite{Mvondo-She:2021joh} we showed that the log partition function is a $\tau$-function of the KP hierarchy of soliton evolution equations, using the relationship between Schur functions and infinite-dimensional Grassmann manifold studied in \cite{sato1983soliton}. As such, the free energy in (i.e the logarithm of) the log partition function is a solution of the Kadomtsev-Petviashvili (KP) hierarchy that is isomorphic to an infinite-dimensional Grassmann manifold called the Universal (or Sato's) Grassmann manifold, and the evolution of the log partition function is parametrized by the motion of a point on that Grassmannian.

In this note, we give a precise identification of the palindromic numerators in two variables of bigraded symmetric orbifold Hilbert series take the form of sums of products of Kostka-Foulkes polynomials associated with a pair of partition $\lambda$ and $\mu=(1^n)$. To the best of our knowledge, our work gives a new description of Hall-Littlewood polynomials and Kostka-Foulkes polynomials as palindromic numerators of quotient expansions in an infinite-dimensional Grassmann manifold.

These palindromic polynomials are also shown to be recursive eigenvalues of a differential operator acting on the Hilbert series.
\section{Background}
The aim of this section is to give a succinct expository account of topics that play a central role in our work. Starting with some definitions of symmetric functions, we provide a brief description of the plethystic substitution notation for symmetric functions, which is essential in subsequently introducing Macdonald polynomials, their associated $q,t$-Kostka polynomials, as well as their geometrical interpretation in terms of bigraded $S_n$-representations related to the geometry of the Hilbert scheme of $n$ points in the plane.

\subsection{Symmetric functions}
Let $R$ be a ring (in this paper, we will work with  $R=\mathbb{C}$), and consider the ring of formal power series in infinitely many variables $R\left[\left[x_1, x_2, \ldots\right]\right]$. Let $S_{\infty}$ denote the group of permutations of $\mathbb{N}$, and consider its action on the ring of formal power series by permuting the variables of the formal power
series $f\left(x_1, x_2, \ldots\right) \in R\left[\left[x_1, x_2, \ldots\right]\right]$ such that for $\pi \in S_{\infty}$

\begin{eqnarray}
\pi \cdot f\left(x_1, x_2, \ldots\right)=f\left(x_{\pi(1)}, x_{\pi(2)}, \ldots\right).    
\end{eqnarray}

The function $f$ is called a symmetric function if $\pi \cdot f=f$ for every $\pi \in S_{\infty}$. Furthermore, the collection of symmetric functions $\Lambda_R \subset R\left[\left[x_1, x_2, \ldots\right]\right]$ is a ring that naturally inherits the grading of $R\left[\left[x_1, x_2, \ldots\right]\right]$, and is generated by the set of homogeneous symmetric polynomials of degree $d$ as 

\begin{eqnarray}
\Lambda_R \left( x_1,x_2, \ldots \right) = \bigoplus_d R \left[ \left[ x_1, x_2, \ldots \right] \right]_d^{S_{\infty}}.  
\end{eqnarray}

A few well-known natural bases for the ring of symmetric functions \cite{macdonald1998symmetric,stanley2011enumerative}, all indexed by partitions $\lambda= \left( \lambda_1,\ldots,\lambda_k \right)$ are listed below.

\paragraph{Monomial symmetric functions}
\begin{eqnarray}
m_\lambda = \sum_{\substack{i_a \neq i_b ~ \forall a,b, \\ i_a < i_b ~\text{if}~ \lambda_a = \lambda_b}} x_{i_1}^{\lambda_1} \cdots x_{i_k}^{\lambda_k}.   
\end{eqnarray}

\paragraph{Homogeneous symmetric functions} $h_\lambda = h_{\lambda_1} \cdot h_{\lambda_2} \cdots h_{\lambda_k}$ with 
\begin{eqnarray}
h_d = \sum_{i_1 \leq \cdots \leq i_d} x_{i_1} \cdots x_{i_d}.   
\end{eqnarray} 

\paragraph{Elementary symmetric functions} $e_\lambda = e_{\lambda_1} \cdot e_{\lambda_2} \cdots e_{\lambda_k}$ with 
\begin{eqnarray}
e_d = \sum_{i_1 < \cdots < i_d} x_{i_1} \cdots x_{i_d}.   
\end{eqnarray}

\paragraph{Power sum symmetric functions} $p_\lambda = p_{\lambda_1} \cdot p_{\lambda_2} \cdots p_{\lambda_k}$ with 
\begin{eqnarray}
p_d = x_1^d + x_2^d + \cdots .    
\end{eqnarray}

\paragraph{Schur functions}
\begin{eqnarray}
s_\lambda = \sum_{T \in \text{SSYT} \left( \lambda \right)}  x^{\alpha_1 (T)} \cdot x^{\alpha_2 (T)} \cdots,
\end{eqnarray}

\noindent where SSYT$\left( \lambda \right)$ is the set of all semistandard Young tableaux of shape $\lambda$, $\alpha_i \left( T \right)$ is the
number of $i$'s in $T$ \cite{fulton1997young}.

The irreducible representations of the symmetric group $S_n$ are also indexed by partitions $\lambda \vdash n$ \cite{macdonald1998symmetric,sagan2013symmetric}. Given an $S_n$-module $V$, denote the irreducible $S_n$-module by $V_\lambda$ such that 

\begin{eqnarray}
V=\bigoplus_{\lambda \vdash n} V_\lambda^{\oplus c_\lambda},    
\end{eqnarray}

\noindent where $c_\lambda$ is the multiplicity of $V_\lambda$, and let $S^\vee_n$ be the set of all representations of $S_n$. Then, the Frobenius characteristic of $V$ is the map $\operatorname{Frob}:S^\vee_n \mapsto \Lambda_R \left( x_1,  x_2, \ldots \right)$ defining the natural correspondence between representations of the symmetric group $S_n$ and the Schur symmetric functions, given by

\begin{eqnarray}
\operatorname{Frob} \left( V \right) = \sum_{\lambda \vdash n} c_\lambda s_\lambda(\mathbf{x}).
\end{eqnarray}

\subsection{Plethysm}
The plethystic substitution is an operation on symmetric functions defined as follows. Let $A= A \left( a_1, a_2, \ldots \right) \in \mathbb{Z} \left[ \left[ a_1, a_2, \ldots \right]\right]$ be a formal sum of monomials with integer coefficients in the variables $a_i$. For a power sum symmetric function $p_k= x^k_1 + x^k_2 + \cdots$, the plethystic substitution of $A$ into $p_k$ is the expression

\begin{eqnarray}
p_k \left[ A \right] = A \left( a_1^k, a_2^k, \ldots \right).
\end{eqnarray}

\noindent For a general symmetric function $f$, express $f=f \left( p_1,p_2,\ldots \right)$ as a polynomial in the power sum symmetric functions $p_k$. Then

\begin{eqnarray}
f \left[ A \right] = f \left( p_1 \left[ A \right], p_2 \left[ A \right], \ldots \right). 
\end{eqnarray}

\noindent When dealing with the plethystic notation, it is often convenient to define

\begin{eqnarray}
\Omega = \exp \left( \sum_{k=1}^\infty \frac{p_k}{k} \right).    
\end{eqnarray}

\noindent Then, from the operations $p_k \left[ A+B \right]= p_k \left[ A \right] + p_k \left[ B \right]$ and $p_k \left[ -A \right] = -p_k \left[ A \right]$, we have 

\begin{eqnarray}
\Omega \left[ A +B\right]= \Omega \left[ A \right] + \Omega \left[ B\right], \quad \Omega \left[ -A \right]= \frac{1}{\Omega \left[ A \right]}.
\end{eqnarray}

\noindent From this and the single-variable evaluation $\Omega \left[ x \right] = \exp \left( \sum_{k \geq 1} x^k/k \right)= 1/(1-x)$, one obtains 

\begin{subequations}
\begin{align}
 \Omega \left[ X \right] &= \prod_i  \frac{1}{1-x_i}= \sum_{n=0}^\infty h_n (x),\\
 \Omega \left[ -X \right] &= \prod_i \left( 1-x_i \right) = \sum_{n=0}^\infty (-1)^n e_n (x). 
\end{align}    
\end{subequations}

 As we will see below, the plethystic substitutions $X \rightarrow X(1-t)$ and $X \rightarrow X/(1-t)$ have important representation-theoretical interpretations in the theory of Macdonald polynomials.  
\subsection{Macdonald symmetric functions, geometry of configuration of $n$ points in the plane, and the log partition function}

\subsubsection{Macdonald polynomials and (q,t)-Kostka Macdonald coefficients}
A remarkable new basis for the space of symmetric functions known as the Macdonald polynomials was introduced in \cite{macdonald1988new}. The elements of this basis are denoted $P_\lambda \left( X; q,t \right)$ where $\lambda$ is a partition, $X$ is the infinite set of variables $x_1,x_2, \ldots$, and $q,t$ are parameters. Their importance stems in part from the fact that by suitable substitutions for the parameters $q$ and $t$, they specialize to many of the well-known bases for the symmetric functions, and one can recover in this manner the Schur functions, the Hall-Littlewood symmetric functions, the Jack symmetric functions, the zonal symmetric functions, the zonal spherical functions, and the elementary and monomial symmetric functions \cite{macdonald1998symmetric}.

Given a partition $\lambda$, we define the Young diagram $\mathcal{Y}(\lambda)$ of the partition as the array of lattice squares $\mathcal{Y}(\lambda) = \left\{ (i,j) \in \mathbb{N} \times \mathbb{N} : j < \lambda_{i+1}  \right\}$, and represent it using the French convention as follows. Given that the parts of $\lambda$ are $\lambda_1 \geq \lambda_2 \geq \cdots \geq \lambda_k > 0$, we let the corresponding Young diagram have $\lambda_i$ lattice squares in the $i^{th}$ row (counting from the bottom up), with $n_\lambda = \sum_{i=1}^k (i-1) \lambda_i$, and adopt the Macdonald convention \cite{macdonald1998symmetric} of calling the $arm$, $leg$, $coarm$ and $coleg$ of a lattice square $\textcolor{red}{s}$ the parameters $a_\lambda $, $l_\lambda $, $a'_\lambda $ and $l'_\lambda $ giving the number of cells of $\lambda$ that are East, North, West and South of $\textcolor{red}{s}$, respectively. In this setting, $n_\lambda = \sum_{s \in \mathcal{Y}(\lambda)} l(s)$.

\begin{figure}[h]
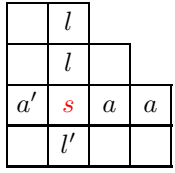

\begin{center}
\ytableausetup{centertableaux}
\begin{ytableau}
\phantom{a} & l & \none & \none \\
  & l &  & \none \\
 a' & \textcolor{red}{s} & a & a \\
  & l' &  & 
\end{ytableau} 
\end{center}
\caption{Young diagram of a partition}
\label{fig1}
\end{figure}

\noindent For example, when $\lambda = (4,4,3,2)$ and \textcolor{red}{s} is the cell with "coordinates" (1,2) pictured in the Young diagram in Fig. (\ref{fig1}), then $a=l=2$, and $a'=l'=1$. The numerical statistic $n_\lambda$ for $\lambda = (4,4,3,2)$ is $n_{(4,4,3,2)}=16$.

The Macdonald polynomials are a class of elements of the ring $\Lambda_{\mathbb{Q} \left( q,t \right)} \left( x_1,x_2, \cdots \right)$ of symmetric functions in the variables $x_i$ with coefficients in the field $\mathbb{Q} \left( q,t \right)$ of rational functions in $q$ and $t$ with rational coefficients, originally defined as the eigenvectors $P_\lambda \left( X;q,t \right) \in \Lambda_{\mathbb{Q} \left( q,t \right)} \left( X \right)$ of certain operators, and expressed as

\begin{eqnarray}
P_\lambda = \sum_{\mu \leq \lambda} a_{\lambda \mu} \left( q,t \right) m_\mu, \hspace{1cm} \text{where} \quad a_{\lambda \lambda}=1.   
\end{eqnarray}

\noindent Macdonald defined the integral form of $P_\lambda$, written $J_\lambda$ to be a $q,t$-scalar multiple of $P_\lambda$, and conjectured that its coefficients with respect to certain other bases are polynomials
in $q$ and $t$, rather than simply being rational functions. Defining

\begin{eqnarray}
J_\lambda \left( X;q,t \right) = c_\lambda \left( q,t \right) P_\lambda \left( X;q,t \right),    
\end{eqnarray}

\noindent where

\begin{eqnarray}
c_\lambda \left( q,t \right) = \prod_{s \in \lambda} \left( 1- q^{a(s)} t^{l(s) +1} \right),  
\end{eqnarray}

\noindent the coefficients $K_{\lambda \mu} \left( q,t \right)$ of $J_\mu$ can be defined in terms of the basis $s_\lambda \left[ X(1-t) \right]$, so that 

\begin{eqnarray}
J_\mu \left( X;q,t \right) = \sum_\lambda K_{\lambda \mu} \left( q,t \right) s_\lambda \left[ X(1-t) \right]. 
\end{eqnarray}

\subsubsection{Geometry of configuration of $n$ points in the plane}
Macdonald conjectured the $K_{\lambda \mu} \left( q,t \right)$ to be polynomials in $q,t$ with positive integer coefficients. In an attempt to prove Macdonald’s conjecture, the so-called \textit{modified Macdonald polynomials} were introduced in \cite{garsia1993graded} as follows. Let 

\begin{eqnarray}
H_\mu \left( X;q, t \right) =  J_\mu \left[ \frac{X}{1-t}; q,t  \right]   
\end{eqnarray}

\noindent be the transformed Macdonald polynomials. There exits a (modified) Macdonald polynomial for every partition $\mu$, written $\Tilde{H}_\mu \left( X; q,t \right)$, and expressed in terms of $H_\mu$ as

\begin{eqnarray}
\Tilde{H}_\mu = t^{n \left( \mu \right)} H_\mu \left( X; q,t^{-1} \right).   
\end{eqnarray}

\noindent The modified $q,t$-Kostka polynomials are then defined as $\Tilde{K}_{\lambda \mu} \left( q,t \right)= t^{n \left( \mu \right)} K_{\lambda \mu} \left( q,t^{-1} \right)$, such that the modified Macdonald polynomials are defined as a two-parameter deformation of the Schur functions $s_\lambda$, and expressed as 

\begin{eqnarray}
\Tilde{H}_\mu \left( X; q,t \right) = \sum_\lambda \Tilde{K}_{\lambda \mu} \left( q,t \right) s_\lambda.   
\end{eqnarray}

\noindent The remarkable connection between the two-parameter symmetric polynomials discovered in \cite{macdonald1988new}, and the geometry of configurations of $n$ points in the affine plane $\mathbb{C}^2$ can be summarized as follows. In \cite{garsia1993graded}, the authors proposed what became known as the $n!$ conjecture, which states that the transformed Macdonald polynomials $\Tilde{H}_\mu \left( X; q,t \right)$ could be realized as the bigraded characters of certain modules for the diagonal action of $S_n$ on two sets of variables. The $n!$ conjecture was proved in \cite{haiman2001hilbert}, consequently establishing Macdonald positivity, i.e $K_{\lambda \mu} \left( q,t \right) \in \mathbb{N} [q,t]$, by connecting a conjectural representation-theoretic interpretation of the transformed $q,t$-Kostka coefficients to the algebraic geometry of the Hilbert scheme of $n$ points in the plane, $\text{Hilb}^n \left( \mathbb{C}^2 \right)$, which can be described as the algebraic variety parametrizing ideals $I \subseteq \mathbb{C} \left[ x,y \right]$, where $\mathbb{C} \left[ x,y \right]$ denotes the coordinate ring of the affine plane, such that dim$_\mathbb{C} \mathbb{C} \left[ x,y \right] / I = n$ (i.e the quotient $\mathbb{C} \left[ x,y \right] / I$ has dimension $n$ as a vector space over $\mathbb{C}$), or equivalently 0-dimensional subschemes of length $n$ in $\mathbb{C}^2$ \cite{haiman2002vanishing,haiman2002combinatorics}.

A singular geometry arises from the Hilbert scheme of $n$ points through the map $ H_n \mapsto \mathbb{C}^{2n}/S_n$, called the Hilbert-Chow morphism. The quotient singularity $\mathbb{C}^{2n}/S_n$ is the geometry of configuration of $n$ points of interest for us.

\subsubsection{The log partition}
In \cite{Mvondo-She:2019vbx}, we showed that the moduli space of the log sector of log gravity is the orbit variety $\left( \mathbb{C}^2 \right)^n / S_n$, where $S_n$ acts on $\left( \mathbb{C}^2 \right)^n$ by permuting coordinates. A standard procedure when discussing $S^n \left( \mathbb{C}^2 \right)$ is to start by considering ordered $n$-tuples of points in the plane, which we can denote by $P_1, \ldots, P_n \in \mathbb{C}^2$. Assigning the points coordinates $x_1,y_1, \ldots x_n, y_n$, the space of all $n$-tuples $\left( P_1, \ldots, P_n \right)$ is identified with $\mathbb{C}^{2n}$. The coordinate ring of $\mathbb{C}^{2n}$ is then the polynomial ring $\mathbb{C} \left[ \bf{x,y} \right]= \mathbb{C} \left[ x_1,y_1, \ldots, x_n, y_n \right]$ in $2n$ variables. It follows from the action of the symmetric group $S_n$ on $\mathbb{C}^{2n}$ that the subring of invariants $\mathbb{C} \left[ \bf{x,y} \right]^{S_n}= \mathbb{C} \left[ x_1,y_1, \ldots, x_n, y_n \right]^{S_n}$ is a direct summand of $\mathbb{C} \left[ \bf{x,y} \right]$ as an $S_n$-module, and also a $\mathbb{C} \left[ \bf{x,y} \right]^{S_n}$-module. Furthermore, the Hilbert series of $\mathbb{C} \left[ \bf{x,y} \right]^{S_n}$ can be computed as a doubly graded ring, by degree in the $\bf{x}$ and $\bf{y}$ variables separately. 

The partition function of the log sector was found to be the generating function with parameters $q$ and $\bar{q}$ 

\begin{eqnarray}
Z_{log}\left( \textcolor{red}{q^2} ;q,\bar{q};\mathbb{C}^2 \right) = 1 + \sum_{n=1}^{\infty} Z_n \left( q,\bar{q};\mathbb{C}^2 \right) \left( \textcolor{red}{q^2} \right)^n,
\end{eqnarray}

\noindent with the Hilbert series of the ring of invariants $\mathbb{C} \left[ \bf{x,y} \right]^{S_n}$ 

\begin{eqnarray}
\label{palindromic Zn}
Z_n \left( q,\bar{q};\mathbb{C}^2 \right) = \frac{P_{m,\bar{m}}( q, \bar{q})}{\prod\limits_{i=1}^n \left( 1-q^i \right) \left( 1-\bar{q}^i \right)},  
\end{eqnarray}

\noindent whose numerator $P_{m,\bar{m}}( q, \bar{q})$ is palindromic, i.e it can be written in the form of a degree $m,\bar{m}$ polynomial in $q,\bar{q}$

\begin{eqnarray}
P_{m,\bar{m}}( q, \bar{q})= \sum_{k=0}^{m} \sum_{\bar{k}=0}^{\bar{m}} a_{k,\bar{k}} q^k \bar{q}^{\bar{k}} ,
\end{eqnarray}

\noindent with symmetric coefficients $a_{m-k,\bar{m}-\bar{k}}=a_{k,\bar{k}}$. 

In addition to the expression (\ref{palindromic Zn}) formulated in \cite{Mvondo-She:2019vbx}, following Macdonald and considering Young diagrams $\mathcal{Y}$ of size $n$, the partition function of the log sector can also be expressed as\footnote{Similar expressions have appeared for instance in Eq. (4.5) of \cite{Nakajima:2003pg}, \cite{macdonald1998symmetric} or \cite{haiman2002vanishing}.}. 

\begin{eqnarray}
Z_{log}\left( \textcolor{red}{q^2} ;q,\bar{q};\mathbb{C}^2 \right) =  \sum_{|\mathcal{Y}|} \frac{\left(\textcolor{red}{q^2}\right)^{|\mathcal{Y}|}}{\prod_{s \in \mathcal{Y}}\left(1-q^{-l(s)} \bar{q}^{1+a(s)}\right)\left(1-q^{1+l(s)} \bar{q}^{-a(s)}\right)}.
\end{eqnarray}
\section{Palindromic numerators of bigraded symmetric orbifold Hilbert series and Kostka-Foulkes polynomials}
In this section, we want to show that the palindromic numerators in two variables of bigraded symmetric orbifold Hilbert series belong to a class of functions called harmonic polynomials, that take the form of sums of products of Kostka-Foulkes polynomials associated with a pair of partition $\lambda$ and $\mu=(1^n)$.

Denoted $K_{\lambda \mu}(q)$, the Kostka-Foulkes polynomials appear in the classical theory of Hall-Littlewood polynomials \cite{macdonald1998symmetric}, which are symmetric functions with coefficients depending on a parameter $q$ generalized by the $q,t$ Macdonald polynomials, and are a specialization of their bivariate analogs, the $q,t$ Kostka-Macdonald polynomials $K_{\lambda\mu}(q,t)$. Since the goal of this section involves  applications of  Schur functions and (transformed) Hall-Littlewood polynomials, we briefly introduce these combinatorial objects before presenting our main result. 

\subsection{Hall-Littlewood polynomials}
Implicitly introduced in terms of the Hall algebra by Philip Hall and later explicitly defined by D.E. Littlewood, the Hall-Littlewood polynomials provide a generalization for the complete homogeneous symmetric functions as well as the Schur symmetric functions, which will both play a role in our interpretation of the palindromic numerators in terms of the Kostka-Foulkes polynomials. The theory of Hall-Littlewood polynomials and its many connections with algebra, geometry and combinatorics can be found in \cite{macdonald1998symmetric}. In Chapter II for instance, the theory describes the structure of the Hall algebra of finite $\mathfrak{o}$-modules with discrete valuation ring $\mathfrak{o}$, and Chapter VI explains the important role the Hall-Littlewood polynomials play in the representation theory of $G L_n\left(\mathbb{F}_q\right)$, where $\mathbb{F}_q$ is a finite field with $q$ elements.

Beginning with the classical definition of the Hall-Littlewood polynomials \cite{macdonald1998symmetric}, for a positive integer $k$, define its $t$-analog and $t$-factorial as

\begin{subequations}
\begin{align}
& {[k]_t = \frac{1-t^k}{1-t}=t^{k-1}+t^{k-2}+\cdots+1}, \\
& {[k]_{t}! =[k]_t[k-1]_t \cdots[1]_t}.
\end{align}
\end{subequations}

\noindent The Hall-Littlewood polynomial $P_\lambda(X ; t)$ is defined in $n \geq l(\lambda)$ variables $X=x_1, \ldots, x_n$ by the formula

\begin{eqnarray}
P_\lambda(X ; t)=\frac{1}{\prod_{i \geq 0}\left[\alpha_i\right]_{t}!} \sum_{w \in S_n} w\left(x^\lambda \frac{\prod_{i<j}\left(1-t x_j / x_i\right)}{\prod_{i<j}\left(1-x_j / x_i\right)}\right),
\end{eqnarray}

\noindent where $\lambda=\left(1^{\alpha_1}, 2^{\alpha_2}, \ldots\right)$, with $\alpha_0$ defined so as to make $\sum_i \alpha_i=n$, and $x^\lambda$ is shorthand for $x_1^{\lambda_1} x_2^{\lambda_2} \cdots x_l^{\lambda_l}$. In the above definition, note that the Hall-Littlewood polynomials generalize the monomial symmetric functions by letting $t \rightarrow 1$, and the Schur symmetric functions by letting $t \rightarrow 0$.

The Hall-Littlewood polynomials $P_\lambda(X ; t)$ may be transformed into $H_\lambda(X ; t)$ by first setting 

\begin{eqnarray}
Q_\mu(X ; t)=(1-t)^{\ell(\mu)} P_\mu(X ; t) \prod_i \left[\alpha_i\right]_{t}!,    
\end{eqnarray}

\noindent and then writing 

$$
Q_\lambda[X ; t]:=H_\lambda \left[(1-t) X ; t\right],
$$

\noindent such that 

$$
H_\lambda[X ; t]=Q_\lambda\left[\frac{X}{(1-t)} ; t\right].
$$

\noindent The transformed Macdonald polynomials $H_\lambda(X ; q, t)$ generalizes the transformed Hall-Littlewood polynomials as, by setting $q=0$, $H_\mu(X ; 0, t)=$ $H_\mu( X; t)$.

The transformed Hall-Littlewood symmetric polynomials can be expanded in terms of the Schur symmetric function basis. For partitions $\lambda$ and $\mu$ of $n$, with $\mu \leq \lambda$ this expansion is given by

\begin{eqnarray}
H_\lambda(X ; t)=\sum_\lambda K_{\lambda \mu}(t) s_\mu(x),    
\end{eqnarray}

\noindent where the coefficients $K_{\lambda \mu}(t)$ are the Kostka-Foulkes polynomials in variable $t$ with non-negative integer coefficients, defined as the matrix elements of the transition matrix from the Schur functions $s_\lambda(X)$ to the Hall-Littlewood polynomials $P_\mu (X ; t)$ in the expansion

\begin{eqnarray}
s_\lambda(X)=\sum_\mu K_{\lambda \mu}(t) P_\mu(X ; t).
\end{eqnarray}

\noindent The conjecture on the positivity of the coefficients in the Kostka-Foulkes polynomials \cite{foulkes1974survey} was the prelude of Macdonald’s positivity conjecture.

\subsection{Harmonic polynomials of the symmetric group}
The transformed Hall-Littlewood polynomials $H_\lambda(X ; t)$ can also be expressed as graded characteristics of the representations of the symmetric group in certain spaces of polynomials \cite{desarmenien1994hall}. Recall that as a bijection between symmetric functions and characters of representations of the symmetric group, the Frobenius characteristic is the linear map from the representation ring of characters of symmetric groups $R(S_n)$ to the ring of symmetric functions which sends the irreducible representation indexed by a partition $\lambda$ to the Schur function $s_\lambda$. For any $S_n$-module $V$, denoting $\chi$ the character of a representation $V$, the Frobenius characteristic map of $V$ is 

\begin{eqnarray}
\mathcal{F} (V) = \sum_{\left| \alpha \right|=n} \chi_\alpha \frac{p_\alpha}{z_\alpha}.    
\end{eqnarray}

\noindent For a graded module, $V= \bigoplus_k V_k$, the graded Frobenius characteristic $\mathcal{F}_t (V)$ is the formal power series

\begin{eqnarray}
\label{frobtv}
\mathcal{F}_t (V) = \sum_k t^k \mathcal{F} (V_k),   
\end{eqnarray}

\noindent with coefficient in the ring $\mathbb{C} \left[ x_1, \ldots, x_n \right]^{S_n}$ of symmetric functions. In particular, if we consider the ring of polynomials $V=\mathbb{C} \left[ x_1, \ldots, x_n \right]$, then

\begin{eqnarray}
\mathcal{F}_t (V) = h_n \left[ \frac{X}{1-t} \right].   
\end{eqnarray}

\noindent Modding out the space of polynomials by the ideal generated by $S_n$-invariant polynomials, one obtains another important $S_n$-module, the space $\mathcal{H}_{S_n}$ of $S_n$-harmonic polynomials. The space $\mathcal{H}_{S_n}$ of $S_n$-invariant polynomials $f \left( x_1, \ldots, x_n \right)$ is closed under partial derivatives and contains the Jacobian determinant 

\begin{eqnarray}
\Delta_{S_n} \left( x_1, \ldots, x_n \right)=\operatorname{det} \frac{\partial f_i}{\partial x_j},
\end{eqnarray}

\noindent associated with any specific set $\left\{f_1, \ldots, f_n\right\}$ of basic invariants for $S_n$, such that the identity

\begin{eqnarray}
f \left( \frac{\partial}{\partial x_1}, \ldots, \frac{\partial}{\partial x_n} \right) \Delta_{S_n} \left( x_1, \ldots, x_n \right) = 0    
\end{eqnarray}

\noindent is satisfied. The link between the companion ring $\mathcal{H}_{S_n}$ to the ring of $S_n$-invariants is established through a result of Chevalley (valid for any reflection group) \cite{bergeron2009algebraic}, which states that there is a natural isomorphism of graded $S_n$-modules

\begin{eqnarray}
\label{isomorphism}
\mathbb{C} \left[ x_1, \ldots, x_n \right] \simeq \mathbb{C} \left[ x_1, \ldots, x_n \right]^{S_n} \otimes \mathcal{H}_{S_n},    
\end{eqnarray}

\noindent where both $\mathbb{C} \left[ x_1, \ldots, x_n \right]^{S_n}$ and $\mathcal{H}_{S_n}$ are $S_n$-submodules of $\mathbb{C} \left[ x_1, \ldots, x_n \right]$.

\noindent Since it is known that the graded enumeration of $V^{S_n}$ is given by 

\begin{eqnarray}
\frac{1}{\prod_{i=1}^n\left(1-t^i\right)}, 
\end{eqnarray}

\noindent and given the result in Eq. (\ref{frobtv}), an important result from the graded $S_n$-module isomorphism in Eq. (\ref{isomorphism}) that connects the graded Frobenius characteristic of $\mathcal{H}_{S_n}$ and the transformed Hall-Littlewood polynomials is 

\begin{eqnarray}
\label{frobhl}
\mathcal{F}_t \left( \mathcal{H}_{S_n} \right) = h_n \left[ \frac{X}{1-t} \right] \prod_{k=1}^n \left( 1- t^k \right) = \sum_{ \left| \lambda \right| =n} K_{\lambda \left( 1^n \right)} (t) s_\lambda \left( X \right),    
\end{eqnarray}

\noindent where the right-hand side of Eq. (\ref{frobhl}) is the transformed Hall-Littlewood polynomials.

\subsection{Palindromic numerators as a sum of products of Kostka-Foulkes polynomials}
We are now ready to show that the palindromic numerators of the generating function $Z_{log}\left( \textcolor{red}{q^2} ;q,\bar{q};\mathbb{C}^2 \right)$ of bigraded symmetric orbifold Hilbert series can be expressed as sums of products of Kostka-Foulkes polynomials associated with a pair of partition $\lambda$ and $\mu=(1^n)$. By relabeling the numerators as $\mathcal{P}_n \left( q, \bar{q}   \right)$, we claim that 

\begin{eqnarray}
\label{palindromic Zn bis}
\boxed{\begin{split}
Z_{log}\left( \textcolor{red}{q^2} ;q,\bar{q};\mathbb{C}^2 \right) &= 1 + \sum_{n=1}^{\infty} Z_n \left( q,\bar{q};\mathbb{C}^2 \right) \left( \textcolor{red}{q^2} \right)^n \\ &= 1 + \sum_{n=1}^{\infty}
 \frac{\mathcal{P}_n( q, \bar{q})}{\prod\limits_{i=1}^n \left( 1-q^i \right) \left( 1-\bar{q}^i \right)} \left( \textcolor{red}{q^2} \right)^n, \\\quad \text{with} \quad   \mathcal{P}_n( q, \bar{q}) &=  \sum_{\left| \lambda \right| = n} K_{\lambda, 1^n}(q) K_{\lambda, 1^n}(\bar{q}).
\end{split}}
\end{eqnarray}

\noindent From Eq. (\ref{palindromic Zn bis}), we first write

\begin{eqnarray}
\mathcal{P}_n( q, \bar{q}) = Z_n \left( q,\bar{q};\mathbb{C}^2 \right)  \prod\limits_{i=1}^n \left( 1-q^i \right) \left( 1-\bar{q}^i \right).  
\end{eqnarray}

\noindent Then, we note that $Z_n \left( q,\bar{q};\mathbb{C}^2 \right)$ can be expressed as a complete homogeneous symmetric function using the identification

\begin{eqnarray}
Z_n \left( q,\bar{q};\mathbb{C}^2 \right) = h_n \left[ \frac{1}{(1-q)(1-\bar{q})} \right].   
\end{eqnarray}

\noindent Finally, from the fact that 

\begin{eqnarray}
 h_n \left[ \frac{X}{1-q} \right] \prod_{k=1}^n \left( 1- q^k \right) = \sum_{ \left| \lambda \right| =n} K_{\lambda \left( 1^n \right)} (q) s_\lambda \left( X \right),    
\end{eqnarray}

\noindent according to Eq. (\ref{frobhl}), and using the Schur polynomial specialization \cite{desarmenien1994hall} 

\begin{eqnarray}
s_\lambda\left[\frac{1}{1-\bar{q}}\right] \prod_{k=1}^n\left(1-\bar{q}^k\right)=K_{\lambda\left(1^n\right)}(\bar{q}),    
\end{eqnarray}

\noindent it follows that 

\begin{eqnarray}
h_n\left[\frac{1}{(1-q)(1-\bar{q})}\right] \prod_{k=1}^n\left(1-q^k\right)\left(1-\bar{q}^k\right)=\sum_{|\lambda|=n} K_{\lambda\left(1^n\right)}(q) K_{\lambda\left(1^n\right)}(\bar{q}).    
\end{eqnarray}

\noindent Using the hook-formula \cite{macdonald1998symmetric}

\begin{eqnarray}
K_{\lambda\left(1^n\right)}(q)=\frac{q^{n\left(\lambda^{\prime}\right)}(q ; q)_n}{\prod_{s \in \lambda}\left(1-q^{h(s)}\right)},
\end{eqnarray}

\noindent where $(q ; q)_n=(1-q)\cdots (1-q^n)$, and $h(x)=1+a(x)+l(x)$ denotes the hook-length of the cell $s$ in the diagram of $\mu$, the above result can be verified at lower levels, giving

\begin{subequations}
\begin{align}
\text{For } n=1:& \quad\mathcal{P}_1 \left( q, \bar{q}   \right)=1,\\
\text{For } n=2:& \quad \mathcal{P}_2 \left( q, \bar{q}   \right) = 1 + q\bar{q},\\
\text{For } n=3:& \quad \mathcal{P}_3 \left( q, \bar{q}   \right) = 1 + (q\bar{q})(1+q)(1+\bar{q}) + (q\bar{q})^3= 1 + q^1\bar{q}^1 + q^2\bar{q}^1 +  q^1 \bar{q}^2 +  q^2 \bar{q}^2 +  q^3 \bar{q}^3,
\\
\text{For } n=4:&  \quad \mathcal{P}_4 \left( q, \bar{q} \right) = 1 +(q \bar{q})\left(1+q+q^2\right) \left(1+\bar{q}+\bar{q}^2\right) + (q \bar{q})^2\left(1+q^2\right)\left(1+\bar{q}^2\right)\\ & \hspace{2cm} +
(q \bar{q})^3\left(1+q+q^2\right)\left(1+\bar{q}+\bar{q}^2\right) + (q \bar{q})^6, 
\end{align}    
\end{subequations}

\noindent which is in agreement with the numerators calculated in \cite{Mvondo-She:2019vbx}.

\noindent We end this section by noting that, collecting $q$ and $\bar{q}$ under a rescaling of variables with coordinate sequence $\left( \mathcal{G}_k \right)_{k=1}^n$ as

\begin{eqnarray}
\mathcal{G}_k \left( q,\bar{q} \right)= \mathcal{G}_1 \left( q^k,\bar{q}^k \right) = \frac{1}{\left( 1-q^k \right) \left( 1-\bar{q}^k \right)},    
\end{eqnarray}

\noindent and writing the log partition function with the plethystic exponential notation, as the generating function of the cycle decomposition of any permutation $\pi$ in the symmetric group $S_n$ of all permutations on integrers $1, \ldots , n$

\begin{eqnarray}
 Z_{log} \left( \mathcal{G}_1, \ldots , \mathcal{G}_n  \right) = \text{PE}_{\left( \textcolor{red}{q^2} \right)} \left[ \mathcal{G}_1 \left( q,t \right) \right] = \exp \left(  \sum_{k= 1}^{\infty} \frac{\mathcal{G}_k \left( \textcolor{red}{q^2} \right)^k}{k} \right) = 1 +  \sum_{n=1}^{\infty} \frac{1}{n!} \left( \sum_{\pi \in S_n} \prod_{k=1}^n \mathcal{G}_k^{j_k} \right)\left( \textcolor{red}{q^2} \right)^n,  
\end{eqnarray}

\noindent where the sequence $\left(j\right)_{k=1}^n$ of positive integers $j_k$ called occupation numbers or cycle counts of any partition in the set of all partitions $\mathcal{P}_n$ of $n$ satisfies the constraint $\sum_{k=1}^n k j_k = n$, and the bigraded Hilbert series of the ring of $S_n$-invariants is

\begin{eqnarray}
\label{cycle index}
Z_n \left( \mathcal{G}_1, \ldots , \mathcal{G}_n  \right) = \frac{1}{n!} \left( \sum_{\pi \in S_n} \prod_{k=1}^n \mathcal{G}_k^{j_k} \right),     
\end{eqnarray}

\noindent our palindromic polynomial $\mathcal{P}_n \left( q, \bar{q}   \right)$ can be expressed in determinantal form as 

\begin{eqnarray}
\label{determinant}
\boxed{
\begin{split}
\mathcal{P}_n \left( q, \bar{q}   \right) &=   \sum_{\left| \lambda \right| = n} K_{\lambda, 1^n}(q) K_{\lambda, 1^n}(\bar{q}) = \frac{Z_n \left( \mathcal{G}_1, \ldots , \mathcal{G}_n  \right)}{\prod\limits_{k=1}^n \mathcal{G}_k \left( q,\bar{q} \right)} \\ &= \frac{1}{n! \prod\limits_{k=1}^n \mathcal{G}_k \left( q,\bar{q} \right)} 
\begin{vmatrix}
\mathcal{G}_1 & \mathcal{G}_2 & \mathcal{G}_3 & \mathcal{G}_4 & \cdots & \cdots & \mathcal{G}_n\\
-1 & \mathcal{G}_1 & \mathcal{G}_2 & \mathcal{G}_3 & \cdots & \cdots & \mathcal{G}_{n-1} \\
0 & -2 & \mathcal{G}_1 & \mathcal{G}_2 & \cdots & \cdots &\mathcal{G}_{n-2}\\
0 & 0 & -3 & \mathcal{G}_1 & \cdots & \cdots & \mathcal{G}_{n-3} \\
0 & 0 & 0 & -4 & \cdots & \cdots & \mathcal{G}_{n-4} \\
\vdots & \vdots & \vdots & \vdots & \ddots & \ddots & \vdots\\
0 & 0 & 0 & 0 & \cdots & -(n-1) & \mathcal{G}_1
\end{vmatrix}.   
\end{split}}
\end{eqnarray}

\noindent In Eq. (\ref{determinant}), we have used the well-known fact that the cycle index $Z_n \left( \mathcal{G}_1, \ldots , \mathcal{G}_n  \right)$ of the symmetric group $S_n$ in Eq. (\ref{cycle index}) can be stated in terms of Bell polynomials, which can be expressed as determinants. We note that the determinant in Eq. (\ref{determinant}) is essentially the determinant for $h_{n}$ in terms of power-sums (see page 28 in \cite{macdonald1998symmetric}).

\section{Recurrences relations and differential operators}
An inductive proof that Eq. (\ref{determinant}) holds can be given by noticing that 

\begin{eqnarray}
\frac{\mathcal{P}_2 \left( q, \bar{q}   \right)}{\mathcal{P}_1 \left( q, \bar{q}   \right)} = \frac{1}{\mathcal{G}_2\left( q,\bar{q} \right)} \frac{Z_2 \left( \mathcal{G}_1, \mathcal{G}_2  \right)}{Z_1 \left( \mathcal{G}_1 \right)}, \hspace{1cm}  \frac{\mathcal{P}_3 \left( q, \bar{q}   \right)}{\mathcal{P}_2 \left( q, \bar{q}   \right)} = \frac{1}{\mathcal{G}_3\left( q,\bar{q} \right)} \frac{Z_3 \left( \mathcal{G}_1, \mathcal{G}_2, \mathcal{G}_3  \right)}{Z_2 \left( \mathcal{G}_1, \mathcal{G}_2  \right)}, \hspace{1cm}  \cdots,
\end{eqnarray}

\noindent from which, the following recurrence relation can be drawn

\begin{eqnarray}
\label{recurrence}    
\mathcal{G}_{n+1} \left( q,\bar{q} \right) \cdot \mathcal{P}_{n+1} \left( q, \bar{q}   \right) \cdot Z_n \left( \mathcal{G}_1, \ldots , \mathcal{G}_n  \right) =  \mathcal{P}_n \left( q, \bar{q}   \right) \cdot Z_{n+1} \left( \mathcal{G}_1, \ldots , \mathcal{G}_{n+1}  \right). 
\end{eqnarray}

\noindent Then, from the recurrence relation (\ref{recurrence}), we write

\begin{eqnarray}
\label{recurrence 2}
\mathcal{P}_{n+1} \left( q, \bar{q}   \right)= \frac{1}{\mathcal{G}_{n+1} \left( q, \bar{q} \right)} \cdot \frac{Z_{n+1} \left( \mathcal{G}_1, \ldots , \mathcal{G}_{n+1}  \right)}{Z_n \left( \mathcal{G}_1, \ldots , \mathcal{G}_n  \right)} \cdot \mathcal{P}_n \left( q, \bar{q}   \right), 
\end{eqnarray}

\noindent and substituting the last term on the right-hand side of Eq. (\ref{recurrence 2}) yields

\begin{subequations}
\begin{align}
\mathcal{P}_{n+1} \left( q, \bar{q}   \right)&= \frac{1}{\mathcal{G}_{n+1} \left( q,\bar{q} \right)} \cdot \frac{Z_{n+1} \left( \mathcal{G}_1, \ldots , \mathcal{G}_{n+1}  \right)}{Z_n \left( \mathcal{G}_1, \ldots , \mathcal{G}_n  \right)} \cdot \frac{Z_n \left( \mathcal{G}_1, \ldots , \mathcal{G}_n  \right)}{\prod\limits_{k=1}^n \mathcal{G}_k \left( q,\bar{q} \right)} \\
&= \frac{Z_{n+1} \left( \mathcal{G}_1, \ldots , \mathcal{G}_{n+1} \right)}{\mathcal{G}_{n+1} \left( q,\bar{q} \right) \prod\limits_{k=1}^n \mathcal{G}_k \left( q,\bar{q} \right)} \\
&= \frac{Z_{n+1} \left( \mathcal{G}_1, \ldots , \mathcal{G}_{n+1}  \right)}{\prod\limits_{k=1}^{n+1} \mathcal{G}_k \left( q,\bar{q} \right)},
\end{align}    
\end{subequations}

\noindent which completes the proof of Eq. (\ref{determinant}).

\noindent We further notice that using the operator 

\begin{eqnarray}
{\rm{\hat{X}}} = \mathcal{G}_1 + \sum_{n=1}^\infty n  \mathcal{G}_{n+1} \frac{\partial}{\partial \mathcal{G}_n},   
\end{eqnarray}

\noindent introduced in \cite{Mvondo-She:2018htn} (also utilized in \cite{Mvondo-She:2023ppz} in the context of urn models) as a multiplication operator\footnote{We note that the operator $\hat{\mathrm{X}}$ when acting on power sums is the special case $\alpha=1$ of the operator on Jack polynomials found in Example 6, on page 385 of \cite{macdonald1998symmetric}.} which together with the lowering operator $\hat{D} = \frac{\partial}{\partial \mathcal{G}_1}$ generates the Heisenberg–Weyl algebra $\left[ \hat{D}, \hat{X} \right]=1$, and whose action raises the partition function $Z_n \left( \mathcal{G}_1, \ldots , \mathcal{G}_n  \right)$ to $Z_{n+1} \left( \mathcal{G}_1, \ldots , \mathcal{G}_{n+1}  \right)$ as

\begin{eqnarray}
{\rm{\hat{X}}} Z_n = Z_{n+1},    
\end{eqnarray}

\noindent Eq. (\ref{recurrence}) can be rewritten as

\begin{eqnarray}
\boxed{\mathcal{P}_{n+1} \left( q, \bar{q}   \right) \cdot Z_n \left( \mathcal{G}_1, \ldots , \mathcal{G}_n  \right) = \left[ \frac{\mathcal{P}_n \left( q, \bar{q}   \right)}{ \mathcal{G}_{n+1}} {\rm{\hat{X}}} \right] \cdot Z_n \left( \mathcal{G}_1, \ldots , \mathcal{G}_n  \right)},   
\end{eqnarray}

\noindent where the operator 

\begin{eqnarray}
\mathcal{D} =  \frac{\mathcal{P}_n \left( q, \bar{q}   \right)}{ \mathcal{G}_{n+1}} {\rm{\hat{X}}},  
\end{eqnarray}

\noindent acts on $Z_n \left( \mathcal{G}_1, \ldots , \mathcal{G}_n  \right)$ giving $\mathcal{P}_{n+1} \left( q, \bar{q}   \right)$ as a polynomial eigenvalue.

\section{Summary and outlook}
We have shown that the palindromic numerators in two variables of bigraded symmetric orbifold Hilbert series take the form of sums of products of Kostka-Foulkes polynomials associated with a pair of partition $\lambda$ and $\mu=(1^n)$. To the best of our knowledge, our work gives a new description of Hall-Littlewood polynomials and Kostka-Foulkes polynomials as palindromic numerators of quotient expansions in an infinite-dimensional Grassmann manifold. Hence our presentation appears as novel. The palindromic polynomials were also shown to be recursive eigenvalues of a differential operator acting on the Hilbert series.

We believe that from the results obtained in this article, an interesting story could emerge in our program of developing phenomenological applications of log gravity and ultimately understanding the AdS$_3$/LCFT$_2$ correspondence. Indeed, some light has been shed on the nature of what could be considered as $n$-particle highest weight states in the log sector, by proving their realization as (products of) harmonic polynomials. Furthermore, it appeared from \cite{Mvondo-She:2024iop} that the log sector shows a remarkable dynamical self-organization into a solitonic configuration that forms a KP hierarchy. The sandpile model being a paradigmatic example of self-organized criticality, we proposed to study an analogy between sandpiles, which for various boundary conditions correspond to distinct LCFT descriptions, and log gravity. Harmonic polynomials have been shown to induce and to provide a framework to analyze sandpile dynamics \cite{lang2019harmonic}, and the results of this paper warrant further study in describing the algebraic structure of the log sector as a self-organized dynamical system. 

\paragraph{Acknowledgements} We thank the anonymous referees for various inputs that contributed to improving the presentation of this paper. This work is supported by the South African Research Chairs initiative of the Department of Science and Technology and the National Research Foundation, and by the National Institute for Theoretical and Computational Sciences, NRF Grant Number 65212.

\clearpage

\bibliographystyle{utphys}
\bibliography{sample}
\end{document}